\documentclass[12pt,preprint]{aastex}
\begin{document}
\title{Polarization Evolution of Early Optical Afterglows of Gamma-Ray Bursts}
\author{Mi-Xiang Lan$^{1,2}$, Xue-Feng Wu$^{3,4}$, and Zi-Gao Dai$^{1,2}$}
\affil{$^{1}$School of Astronomy and Space Science, Nanjing University, Nanjing 210093, China; dzg@nju.edu.cn \\
$^{2}$Key Laboratory of Modern Astronomy and Astrophysics (Nanjing
University), Ministry of Education, China \\
$^{3}$Purple Mountain Observatory, Chinese Academy of Sciences,
Nanjing 210008, China\\
$^{4}$Joint Center for Particle Nuclear Physics and Cosmology of
Purple Mountain Observatory-Nanjing University, Chinese Academy of
Sciences, Nanjing 210008, China}

\begin{abstract}
The central engine and jet composition of gamma-ray bursts (GRBs) remain mysterious. Here we suggest that observations on polarization evolution of early optical afterglows may shed light on these questions. We first study the dynamics of a reverse shock and a forward shock that are generated during the interaction of a relativistic jet and its ambient medium. The jet is likely magnetized with a globally large-scale magnetic field from the central engine. The existence of the reverse shock requires that the magnetization degree of the jet should not be high ($\sigma\leq 1$), so that the jet is mainly composed of baryons and leptons. We then calculate the light curves and polarization evolution of early optical afterglows, and find that when the polarization position angle changes by $90^\circ$ during the early afterglow, the polarization degree is zero for a toroidal magnetic field but is very likely to be non-zero for an aligned magnetic field. This result would be expected to provide a probe for the central engine of GRBs, because an aligned field configuration could originate from a magnetar central engine and a toroidal field configuration could be produced from a black hole via the Blandford-Znajek mechanism. Finally, for such two kinds of magnetic field configurations, we fit the observed data of the early optical afterglow of GRB 120308A equally well.
\end{abstract}


\section{Introduction}
Gamma-ray bursts (GRBs) are the most violent explosive events occurring at the cosmological distances. After their prompt emission, a relativistic jet interacts with its ambient medium, leading to two shocks: a reverse shock that propagates into the jet and a forward shock that propagates into the medium. Afterglows are thought to be produced from these shocks (for reviews see Piran 1999; van Paradijs et al. 2000; M\'esz\'aros 2002; Zhang \& M\'esz\'aros 2004). Since the first optical afterglow was discovered in GRB 970228 (Groot et al. 1997; van Paradijs et al. 1997), many optical afterglows have been detected (for a summary see Li et al. 2012; Liang et al. 2013), of which only about ten have been detected with polarized emission (i.e., Covino et al. 1999). Early optical afterglows as well as their polarization evolution are particularly important, because they may provide useful information about the GRB jets and central engines. For example, with polarization observations of early optical afterglows of GRB 090102 (Steele et al. 2009) and GRB 060418 (Mundell et al. 2007), it was suggested that the magnetization degree in the ejecta of these two bursts may range from 0.01 to 0.1 (Kobayashi 2012).

Dynamics which describes the evolution of forward-reverse shocks or a relativistic forward shock has been discussed widely (Blandford \& McKee 1976; Sari \& Piran 1995; Huang et al. 1999, hereafter HDL99; Kobayashi 2000; Pe'er 2012; Nava et al. 2013). At an early stage, assuming equality of the pressure and velocity along a contact discontinuity between the two shocks, the dynamics of a system containing the two shocks can be derived under two extreme conditions, i.e. thick-shell and thin-shell (Sari $\&$ Piran 1995; Kobayashi 2000). At a very late stage, the system enters the Sedov-Taylor evolution phase and its dynamics can be derived from the conservation of the kinetic energy. So there is a transition between these two phases. HDL99 studied the dynamics of a forward shock by considering the conservation of the kinetic energy and proposed a generic dynamical model, which can describe the hydrodynamic evolution from the early ultrarelativistic phase to late non-relativistic phase. Pe'er (2012) developed a slightly-different dynamical model, which includes contribution of the pressure of the shock-heated inter-stellar medium to the total energy of the system. Recently, Nava et al. (2013) considered a time-varying radiative efficiency of the shock and their dynamical equations are valid for an arbitrary density profile, especially for an electron-positron-pair-enriched medium. In this paper, we study the evolution of the system including contributions from forward-reverse shocks by considering conservation of the kinetic energy of the system.

Whether an early optical afterglow is bright or dark depends mainly on the magnetization degree of the jet, $\sigma\equiv L_c/L_h$, where $L_c$ and $L_h$ are the luminosities of the Poynting flux and the kinetic flux, respectively. If the magnetization degree of the jet is very high, i.e. $\sigma\gg1$, the reverse shock emission will be suppressed dramatically (Zhang \& Kobayashi 2005). This may be the reason for many GRBs that do not have very bright early optical afterglow. This implies that the GRB jet with a bright early optical afterglow should be mainly composed of baryons and leptons. In this paper, we show that magnetic field configuration of the jet will affect the polarization evolution significantly. We discuss two magnetic field configurations, i.e. toroidal and aligned (Spruit et al. 2001; Lazzati 2006). In the forward shocked region, because there is no mechanism or process that can produce a large-scale ordered magnetic field, we only consider the random magnetic field in this region and assume that this random field is within the shock plane. The jet structure may also play an important role in the polarization evolution. Here we only consider a homogeneous (top-hat) jet.

This paper is arranged as follows. In Section 2, we present our hydrodynamic model, which extends the generic model of HDL99 by considering the forward-reverse shock interaction. In Section 3, large-scale magnetic field and polarization models of GRB jets in the early afterglow phase are described. In Section 4, we present our numerical results of the light curves and polarization evolution of early afterglows for different hydrodynamics and for different large-scale magnetic field configurations. In Section 5, we apply our model to GRB 120308A, and successfully interpret the light curve and polarization evolution of the early afterglow of GRB 120308A. In Section 6, conclusions and discussion are presented.

\section{Dynamics of Reverse and Forward Shocks}
An ultrarelativistic jet (i.e. ejecta) from the GRB central engine collides with an ambient gas (for simplicity in this work we consider an interstellar medium, ISM), generating two shocks: a reverse shock and a forward shock. Four regions are separated by these two shocks, i.e., the unshocked jet (Region 4), the shocked jet (Region 3), the shocked ISM (Region 2) and the unshocked ISM (Region 1). The total kinetic energy of this system in the burst frame can be expressed by
\begin{equation}
E_{k}=\cases{(\gamma-1)(M_{sw}+M_{rs})c^2+(\eta-1)(M_{ej}-M_{rs})c^2+(1-\varepsilon_{rs})\gamma U'_{rs}+(1-\varepsilon_{fs})\gamma U'_{fs}, & $t < t_c$, \cr (\gamma-1)(M_{sw}+M_{ej})c^2+\gamma [(1-\varepsilon_{rs})U'_{rs}(t_c)+E'_{ad}]+(1-\varepsilon_{fs})\gamma U'_{fs}, & $t \geq t_c$,}
\end{equation}
where $\gamma$ is the bulk Lorentz factor of the shocked jet and also of the shocked ISM (assuming the equality of the velocity along the contact discontinuity between the two shocks), $\eta=E_j/M_{ej}c^2$ is the initial Lorentz factor of the ejecta, $M_{ej}$ and $E_j$ are the initial total (collimation-corrected) mass and energy of the ejecta, respectively, and $c$ is the speed of light. $M_{sw}=2\pi (1-\cos\theta_j)R^3n_1m_p/3$ is the swept-up mass by the forward shock, where $\theta_j$ is the half-opening angle of the jet, $R$ is the radius of the forward shock, $n_1$ is the number density of the ISM, and $m_p$ is the rest mass of proton. $M_{rs}$ is the mass in Region 3, i.e., the mass swept by the reverse shock. We assume that the newly shocked electrons instantaneously radiate a fraction $\varepsilon_{rs}$ for the reverse shock and $\varepsilon_{fs}$ for the forward shock of their internal energy (the cooling of protons is inefficient and can be neglected). Hereafter we define physical quantities in the comoving frame with a prime. $U'_{rs}$ and $U'_{fs}=(\gamma-1)M_{sw}c^2$ are the internal energy of the reverse shock and forward shock, respectively. $t_c$ is the time when the reverse shock has just crossed the ejecta, which corresponds to $M_{rs}=M_{ej}$. When $t < t_c$, $U'_{rs}=(\gamma_{34}-1)M_{rs}c^2$, where $\gamma_{34}$ is the relative Lorentz factor between Region 3 ($\gamma_3$) and Region 4 ($\gamma_4$), which is also the Lorentz factor of the reverse shock measured in Region 4 (upstream). In fact, we have $\gamma_3=\gamma_2=\gamma$ and $\gamma_4=\eta$. $E'_{ad}(t)$ is the energy loss of Region 3 by adiabatic expansion after $t_c$. The width of the reverse shocked region increases by $dX$ when the forward shock propagates radially by $dR$. The relation between these two distances is $dX=(\beta_4-\beta_3)dR/(\gamma_3n'_3/\gamma_4n'_4-1)\beta_{sh}$ (Kobayashi 2000; Yi, Wu \& Dai 2013), where $\beta_3$, $\beta_4$ and $\beta_{sh}$ are the velocities of Region 3, Region 4 and the forward shock, respectively. $n'_3$ and $n'_4=E_j/2\pi (1-\cos\theta_j)R^2m_pc^2\Delta\eta^2$ are the comoving number densities of Region 3 and of Region 4, respectively. Due to the radial spreading, the width of the ejecta increases with time, $\Delta=\Delta_0+c_s t'/\eta$, where $\Delta_0$ is the initial width. Since the comoving spreading speed $c_s\sim c$ and the comoving time $t'\sim \eta t\sim R/\eta c$, the width of the ejecta increases with radius as $\Delta \simeq\Delta_0+R/\eta^2$. In the thick shell case, the initial width of the ejecta is so large that the ejecta does not experience significant radial spreading even at the time the reverse shock crosses the whole ejecta. In this case, the number density of Region 4 decreases with radius as $n'_4=E_j/2\pi (1-\cos\theta_j)R^2m_pc^2\Delta_0\eta^2$. On the other hand, in the thin shell case, the initial width is so small that during the reverse shock crossing the ejecta, the width of the ejecta is dominated by the radial expansion. Therefore in this case $n'_4=E_j/2\pi (1-\cos\theta_j)R^3m_pc^2$. According to the shock jump condition, $n'_3=4\gamma_{34}n'_4$. The shocked mass in Region 3 is $M_{rs}=\int dM_{rs}$ with  $dM_{rs}=2\pi (1-\cos\theta_j)m_p\gamma_3 n'_3R^2dX$.

We do not consider the adiabatic loss because its effect on the hydrodynamic evolution of the system is negligible when the reverse shock is still crossing. The decrease of the total kinetic energy of the system is due to the energy that is radiated away, i.e.,
\begin{equation}
dE_{k}=\cases{-\varepsilon_{rs}\gamma (\gamma_{34}-1)dM_{rs}c^2-\varepsilon_{fs}\gamma (\gamma-1)dM_{sw}c^2, & $t < t_c$, \cr -\varepsilon_{fs}\gamma (\gamma-1)dM_{sw}c^2, & $t \geq t_c$.}
\end{equation}
Therefore, Equation (1) can be easily reduced to $E_k$ of HDL99 with $M_{rs}=M_{ej}$ and $U'_{rs}=0$ before $t_c$ and with $U'_{rs}(t_c)=0$ and $E'_{ad}(t)=0$ after $t_c$, as HDL99 did not take into account the internal energy of the ejecta through the reverse shock heating in their hydrodynamic treatment.

Combing Equation (1) and Equation (2), the hydrodynamics before and after the reverse shock crossing time evolves as
\begin{equation}
\frac{d\gamma}{dR}=-\frac{\displaystyle(\gamma \gamma_{34}-\eta)\frac{dM_{rs}}{dR}+(\gamma^2-1)\frac{dM_{sw}}{dR}}  {(2\gamma-2\varepsilon_{fs}\gamma+\varepsilon_{fs})M_{sw}+[1+(1-\varepsilon_{rs})(\gamma_{34}-1+\gamma \alpha_{34})]M_{rs}}
,\ \ t<t_c,
\end{equation}
\begin{equation}
\frac{d\gamma}{dR}=-\frac{\displaystyle(\gamma^2-1)\frac{dM_{sw}}{dR}+\frac{\gamma}{c^2}\frac{dE'_{ad}}{dR}}{\displaystyle M_{ej}+(1-\varepsilon_{rs})\frac{U'_{rs}(t_c)}{c^2}+\frac{E'_{ad}}{c^2}+(2\gamma-2\varepsilon_{fs}\gamma+\varepsilon_{fs})M_{sw}}
,\ \ t\geq t_c,
\end{equation}
where $\alpha_{34}=d\gamma_{34}/d\gamma$. Before $t_c$, if we let $M_{rs}=M_{ej}$ ($dM_{rs}/dR=0$) and $\gamma_{34}=1$ ($\alpha_{34}=0$), the above Equation (3) is reduced to Equation (7) of HDL99. After $t_c$, if we ignore the internal energy of the reverse shocked region, i.e. $U'_{rs}(t_c)=0$ and $E'_{ad}=0$, Equation (4) is reduced to be exactly consistent with Equation (7) of HDL99.

Relativistic shocks usually have a negligible width of the shocked region compared with the radius (Blandford \& McKee 1976), so we assume Regions 2, 3 and 4 have the same radius which increases with the observer's time as (see also HDL99) due to the relativistic propagation effect,
\begin{equation}
dR=\beta_{sh} c\gamma(\gamma+\sqrt{\gamma^2-1})dt,
\end{equation}

From Equation (96) of Mignone et al. (2005), the internal energy density is $e'=\rho'c^2(5-3\hat{\gamma})/(3\hat{\gamma}-4)$, where $\hat{\gamma}$ is the adiabatic index. As the ejecta evolves from the relativistic phase to the non-relativistic phase, the adiabatic index of the shocked ejecta $\hat{\gamma_3}$ and shocked ISM $\hat{\gamma_2}$ should also change with time, which can be approximated as $\hat{\gamma_2}=4/3+1/3\gamma$ (arbitary $t$) and $\hat{\gamma_3}=4/3+1/3\gamma_{34}$ $(t<t_c)$ (Dai, Huang \& Lu 1999). Note that the residual internal energy in the reverse shocked region after $t_c$ can be expressed as $U'_{rs}(t)=e'_3V'_3=M_{ej}c^2(5-3\hat{\gamma_3})/(3\hat{\gamma_3}-4)$, where $V'_3(t)$ is the comoving volume of Region 3. We can estimate the adiabatic index of the shocked ejecta after $t_c$ by the following equation,
\begin{eqnarray}
\hat{\gamma_3}(t)=\frac{4U'_{rs}(t)+5M_{ej}c^2}{3[U'_{rs}(t)+M_{ej}c^2]}.
\end{eqnarray}
The residual internal energy in Region 3 after $t_c$ can also be expressed as $U'_{rs}(t)=(1-\varepsilon_{rs})U'_{rs}(t_c)+E'_{ad}(t)$, where $U'_{rs}(t_c)$ is the internal energy of Region 3 at time $t_c$, and
\begin{equation}
E'_{ad}(t)=-(\hat{\gamma_3}(t_c)-1)e'_3(t_c)V_3^{'\hat{\gamma_3}(t_c)}(t_c)\int \frac{dV'_3}{V_3^{'\hat{\gamma_3}(t)}(t)}.
\end{equation}
Before $t_c$, the comoving volume of Region 3 is $V'_3(t)=\int dV'_3$ with $dV'_3=2\pi (1-\cos\theta_j)R^2\gamma dX$. After $t_c$, this volume becomes $V'_3(t)=V'_3(t_c)\gamma(t_c)R^3(t)/\gamma(t)R^3(t_c)$.

\section{Magnetic Field Configurations and Polarization}
Early optical flashes of GRBs are widely thought to be generated by a reverse shock. This requires that the magnetization degree of the ejecta can not be high. Otherwise, the emission from the reverse shock will be suppressed seriously (Zhang \& Kobayashi 2005). Initially, the ejecta from the central engine might have a very high magnetization degree ($\sigma\gg1$), depending on the nature of the central engine. During the prompt phase, shells with different velocities within the ejecta collide with each other, so that magnetic fields in the shells will be disturbed and magnetic-reconnection processes might happen (Zhang \& Yan 2011; Deng et al. 2015). After the prompt phase, the magnetization degree of the ejecta may decrease to a moderate level ($\sigma\lesssim1$). Thus, the ejecta in the early afterglow phase may be mainly composed of baryons and leptons and a large-scale, ordered magnetic field remains in the ejecta during the afterglow phase.

Two ordered magnetic field configurations in the ejecta are considered, i.e. toroidal and aligned (Spruit et al. 2001; Lazzati 2006). The random magnetic field generated by the reverse shock in Region 3 is neglected and we assume that in Region 3, the ratio of the energy density of the ordered magnetic field and the internal energy is $\varepsilon_{B,rs}$, which is in fact about $9/2$ of the magnetization parameter $\sigma$  for $\sigma\ll1$. We assume that in the early afterglow phase $\sigma$ does not evolve significantly and keeps as a constant. In our calculation, the value of $\sigma$ is small ($\sigma\sim0.02$ so that $\varepsilon_{B,rs}=0.1$) and the magnetic energy of the ordered field is frozen in Region 3. Therefore, our dynamics discussed in Section 2 is reasonable without considering the effect of the Poynting flux to the jump condition as well as to the total kinetic energy. Since the ordered magnetic field in the ISM may be very weak or even do not exist, we only consider the random magnetic field generated by the forward shock in Region 2. For simplicity, we consider that the random field in the forward shocked region is in the shock plane (for a discussion see Toma et al. 2009).

For GRBs, two possible kinds of central engine are black holes (Narayan, Paczy\'nski, \& Piran 1992; Woosley 1993; M\'esz\'aros \& Rees 1997; Paczy\'nski 1998) and magnetars (Usov 1992; Duncan \& Thompson 1992; Klu\'zniak \& Ruderman 1998; Dai \& Lu 1998a,1998b; Spruit 1999; Ruderman, Tao, \& Klu\'zniak 2000; Wheeler et al. 2000). If the central engine is a black hole, the Blandford-Znajek mechanism would work and a magnetized jet can be powered (Blandford \& Znajek 1977), in which the magnetic field is very likely to be toroidally ordered. For a magnetar central engine, however, the ordered magnetic field in the jet is possibly aligned.

\subsection{Polarization from a Random Magnetic Field}
If the magnetic field is completely random, the net polarization should be zero. If some anisotropy exists, as discussed by Sari (1999), the net polarization will appear. Here we assume that the random magnetic field is in the shock plane which is probably generated by the forward shock. Because the direction of the magnetic field in a point-like region is random, the pitch angle $\theta'_B$ of the electrons is also stochastic. As defined previously, symbols with a prime denote quantities in the comoving frame. From Rybicki $\&$ Lightman (1979), the synchrotron emission power per unit frequency emitted by one single electron in the random magnetic field is
\begin{equation}
p'(\nu')=\frac{\sqrt{3}e^3B'\langle \sin\theta'_B\rangle}{m_ec^2}F\left(\frac{\nu'}{\nu'_c}\right),
\end{equation}
where $e$ and $m_e$ are the charge and rest-frame mass of an electron respectively. $B'$ is the strength of the magnetic field. $\nu'_c=eB'\langle \sin\theta'_B\rangle\gamma_e^2/2\pi m_ec$ is the characteristic frequency of the synchrotron radiation by the electron with Lorentz factor $\gamma_e$ in the random magnetic field. $\nu'=\nu_{obs}(1+z)/D$ is the observational frequency in the comoving frame. $z$ is the redshift of the source and $D=1/\gamma(1-\beta \cos\theta)$ is the Doppler factor. $\gamma$ and $\beta$, as defined in Section 2, are the bulk Lorentz factor and velocity of the flow, respectively. $\theta$ is the angle between the velocity of jet element and the line of sight (LOS) in the observer frame. From Toma et al. (2009), two right-handed coordinates are established. For the coordinate system 123, we set the observational direction $\hat{k}'$ (vector with a hat is a unit vector, same in the following) in the comoving frame to be axis 3. The direction of the magnetic field $\hat{B}'$ is described by its polar and azimuthal angles $\theta'_B$ and $\phi'_B$. Because we assume that the magnetic field is confined in the shock plane, so we set another coordinate system $xyz$ with $z$ being the direction of the velocity of the jet element and $\hat{k}'$ in the $x-z$ plane. The angle between the $z$-axis and $\hat{k}'$ is $\theta'$. So $\sin\theta'=D\sin\theta$. The azimuthal angle of $\hat{B}'$ in the $xyz$ system is $\eta'$. We thus have (Toma et al. 2009)
\begin{equation}
\sin\theta'_B=\left(1-D^2\sin^2\theta\cos^2\eta'\right)^{1/2},
\end{equation}
and
\begin{equation}
\cos(2\phi'_B)=\frac{2\sin^2\eta'}{\sin^2\theta'_B}-1,
\end{equation}
Our Equation (9) and Equation (10) are consistent with Equation (A8) in Toma et al. (2009) under the limit of $\gamma\gg1$ and $\theta\ll1$. We use angle brackets to denote the average over the random magnetic field directions. So $\langle \sin\theta'_B\rangle=\int^{2\pi}_0\sin\theta'_Bd\eta'/2\pi$. The Stokes parameters in coordinate system 123 can be expressed as $q'_{\nu'}=-f'_{\nu'}\pi_0\cos(2\phi'_B)$ and $u'_{\nu'}=-f'_{\nu'}\pi_0\sin(2\phi'_B)$. In order to get the polarization degree from a point-like region, we average the Stokes parameters over the magnetic field direction which leads to $\langle u'_{\nu'}\rangle=0$. The local polarization degree can be expressed as
\begin{equation}
\pi_p(t,\theta)=\frac{\langle q'_{\nu'}\rangle}{\langle f'_{\nu'}\rangle}=-\pi_0\frac{\langle(\sin\theta'_B)^{1-m}\cos(2\phi'_B)\rangle}{\langle(\sin\theta'_B)^{1-m}\rangle},
\end{equation}
where $\pi_0$ is the polarization degree from a smaller region where the magnetic field has a fixed direction. From the conventional notion, the flux density of the electron synchrotron radiation is $f'_{\nu'}\propto (\nu')^m $. The spectrum of synchrotron emission can be well approximated by some power laws (Sari et al. 1998). Different spectral regimes, divided by several synchrotron characteristic frequencies, have different values of $m$. When calculating the polarization evolution of the forward shock emission with a random magnetic field, we use the values of the spectral index $m$ estimated in an analytic way. Given the energy distribution of electrons, the radiation power is
\begin{equation}
P'(\nu')=\int N(\gamma_e)p'(\nu')d\gamma_e.
\end{equation}
where $N(\gamma_e)$ is the energy spectrum of the radiating electrons (Huang $\&$ Cheng 2003). From Huang et al. (2000), the observed flux density of Region 2 is
\begin{equation}
F_{\nu,2}=\frac{1+z}{4\pi D_L^2}\int P'(\nu')D^3\sin\theta d\theta \int d\phi,
\end{equation}
where $D_L$ is the luminosity distance of the source. The position angle $\phi$ for a point-like region in the burst source frame is the angle in the plane of the sky between the projection of the jet axis and the projection of the velocity of the jet element. And $Q_{\nu,2}$ is given by
\begin{equation}
Q_{\nu,2}=\frac{1+z}{4\pi D_L^2}\int P'(\nu')\pi_pD^3\sin\theta d\theta \int \cos2\phi d\phi.
\end{equation}

The integral range of Stokes parameters is over the whole jet, which means that $\theta$ is from 0 to $\theta_j+\theta_V$ and $\phi$ is from $-\Delta\phi$ to $\Delta\phi$. So $U_{\nu,2}\propto \int \sin2\phi d\phi=0$. The $\Delta\phi$ can be expressed by (Wu et al. 2005)
\begin{equation}
\Delta\phi=\left\{ \begin{array}{l}
\displaystyle\pi\Theta(\theta_j-\theta_V), \phantom{ssssssssssssssss} \theta\leq\theta_-,\\
\displaystyle\arccos\left(\frac{\cos\theta_j-\cos\theta_V\cos\theta}{\sin\theta_V\sin\theta}\right), \ \ \theta_-<\theta<\theta_+,\\
\displaystyle0, \phantom{ssssssssssssssssssssssssss}  \theta\geq\theta_+,\\
 \end{array} \right.
\end{equation}
where $\theta_-=|\theta_j-\theta_V|$ and $\theta_+=\theta_j+\theta_V$. $\Theta(x)$ is the Heaviside step function. $\theta_V$ is the observational angle, which is the angle between the jet axis and the LOS.

\subsection{Polarization from a Globally Ordered Magnetic Field}
We assume that in Region 3 the magnetic field is globally ordered and is confined within the shock plane. And we also assume that the ratio of the energy density of the ordered magnetic field and the total energy density generated by the reverse shock in Region 3 is $\varepsilon_{B,r}$. The random magnetic field generated by the reverse shock is neglected. Following the previous work (Granot \& K\"{o}nigl 2003; Granot 2003; Lyutikov et al. 2003; Toma et al. 2009), we discuss the polarization evolution in the early optical afterglow phase.

From Rybicki $\&$ Lightman (1979), the synchrotron emission power per unit frequency emitted by one single electron in the ordered magnetic field is
\begin{equation}
p'(\nu')=\frac{\sqrt{3}e^3B'\sin\theta'_B}{m_ec^2}F\left(\frac{\nu'}{\nu'_c}\right),
\end{equation}
where $\nu'_c=eB'\sin\theta'_B\gamma_e^2/2\pi m_ec$ is the critical frequency of the electron with Lorentz factor $\gamma_e$. The pitch angle $\theta'_B$ in ordered magnetic field configuration can be expressed by

\begin{equation}
\sin\theta'_B=\left[1-D^2\frac{\sin^2\theta\cos^2\varphi}{\cos^2\theta+\sin^2\theta\cos^2\varphi}\right]^{1/2},
\end{equation}
where $\varphi$ is the angle in the plane of the sky between the projection of the ordered magnetic field and the projection of the velocity of the jet element. For $\gamma\gg1$ and $\theta\ll1$, the approximation of the above $\sin\theta'_B$ is consistent with Equation (A2) in Toma et al. (2009).

The observed flux density from Region 3 can be expressed as follows (Huang et al. 2000)
\begin{equation}
F_{\nu,3}=\frac{1+z}{4\pi D_L^2}\int D^3\sin\theta d\theta \int P'(\nu')d\phi,
\end{equation}
where $P'(\nu')=\int N(\gamma_e)p'(\nu')d\gamma_e$. We denote the polarization degree and position angle from a point-like region as $\pi_0$ and $\chi$. Therefore, the Stokes parameters $U_{\nu,3}$ and $Q_{\nu,3}$ can be calculated by
\begin{equation}
U_{\nu,3}=\pi_0\frac{1+z}{4\pi D_L^2}\int D^3\sin\theta d\theta \int P'(\nu')\sin2\chi d\phi,
\end{equation}
and
\begin{equation}
Q_{\nu,3}=\pi_0\frac{1+z}{4\pi D_L^2}\int D^3\sin\theta d\theta \int P'(\nu')\cos2\chi d\phi.
\end{equation}

\subsubsection{Toroidal Magnetic Field}
If a magnetic field is axis-symmetric about the jet axis, it may be toroidal. For this magnetic field configuration, we have
\begin{equation}
\cos\varphi=\frac{\sin\theta_V\cos\theta\sin\phi}{\sqrt{\cos^2\theta_V\sin^2\theta\sin^2\phi+(\sin\theta_V\cos\theta-
\cos\theta_V\sin\theta\cos\phi)^2}}.
\end{equation}
So the pitch angle of electrons in a point-like region in the toroidal magnetic field configuration can be expressed as
\begin{equation}
\sin\theta'_B=\left[1-D^2\frac{\sin^2\theta_V\sin^2\theta\sin^2\phi}{\sin^2\theta\sin^2\phi+(\sin\theta_V\cos\theta-
\cos\theta_V\sin\theta\cos\phi)^2}\right]^{1/2},
\end{equation}
and the position angle for a point-like region can be expressed as
\begin{equation}
\chi=\phi+\arctan\left(\frac{\cos\theta-\beta}{\cos\theta(1-\beta\cos\theta)}\times \frac{\sin\theta_V\cos\theta\sin\phi}{(\cos\theta_V\sin\theta-\sin\theta_V\cos\theta\cos\phi)}\right).
\end{equation}

The above two equations are consistent with Equations (9) and (10) of Toma et al. (2009) under the limit of $\gamma\gg1$ and $\theta\ll1$. Because $P'(\nu')\sin2\chi$ is the odd function with respect to $\phi$, the Stokes parameter $U_{\nu,3}$ is zero when integrating over $\phi$. If the ordered magnetic field in Region 3 is toroidal and the magnetic field in Region 2 is random, the polarization degree of the emission from the forward-reverse shock can be calculated by
\begin{equation}
\Pi_T=\frac{Q_{\nu,2}+Q_{\nu,3}}{F_{\nu,2}+F_{\nu,3}}.
\end{equation}

According to the equations derived above, if one of the Stokes parameters is zero, for example, the polarization degree can be simply expressed as $\Pi=\Pi_T=Q_{\nu}/F_{\nu}$ in the toroidal magnetic field case ($U_{\nu}=0$). Depending on the sign of the Stokes parameter $Q_{\nu}$, the polarization degree $\Pi_T$ can be positive or negative. In this case, the polarization direction for $Q_{\nu}>0$ is perpendicular to that with $Q_{\nu}<0$. In other words, when the polarization degree $\Pi_T$ changes from negative to positive or from positive to negative, the position angle changes abruptly by $90^\circ$.

\subsubsection{Aligned Magnetic Field}
For an aligned magnetic field, let $\delta$ be the orientation of the aligned magnetic field from the projection of the jet axis in the plane of sky. We have $\varphi=\phi-\delta$. The pitch angle of electrons in a point-like region in the aligned magnetic field configuration can be expressed as
\begin{equation}
\sin\theta'_B=\left[1-D^2\frac{\sin^2\theta\cos^2(\phi-\delta)}{\cos^2\theta+\sin^2\theta\cos^2(\phi-\delta)}\right]^{1/2},
\end{equation}
The position angle for a point-like region can be expressed by
\begin{equation}
\chi=\phi+\arctan\left(\frac{\cos\theta-\beta}{\cos\theta(1-\beta\cos\theta)}\cot(\phi-\delta)\right),
\end{equation}

For the aligned magnetic field configuration, generally speaking, both $U_{\nu,3}$ and $Q_{\nu,3}$ are nonzero. The polarization degree and the position angle of the emission from the forward-reverse shock with the aligned magnetic field in Region 3 and the random magnetic field in Region 2 can be calculated by
\begin{equation}
\Pi_A=\frac{\sqrt{(Q_{\nu,2}+Q_{\nu,3})^2+U_{\nu,3}^2}}{F_{\nu,2}+F_{\nu,3}},
\end{equation}
and
\begin{equation}
\chi_A=\frac{1}{2}\arctan\left(\frac{U_{\nu,3}}{Q_{\nu,2}+Q_{\nu,3}}\right).
\end{equation}

If both Stokes parameters $Q_{\nu}$ and $U_{\nu}$ are generally nonzero, as in the aligned magnetic field case, the polarization degree can be expressed as $\Pi=\Pi_A=\sqrt{Q_{\nu}^2+U_{\nu}^2}/F_{\nu}$. By the above definition, $\Pi_A$ is always positive. The position angle is depending on the ratio of the Stokes parameters $Q_{\nu}$ and $U_{\nu}$. Since the sign of $Q_{\nu,2}$ and $Q_{\nu,3}$ may be different, and the dominant contribution to $Q_{\nu}=Q_{\nu,2}+Q_{\nu,3}$ can convert from one to another, then the position angle of the polarization can be changed abruptly by 90$^{\circ}$ (when $Q_{\nu}$ crosses zero) while the degree of polarization changes gradually. This is a unique signature of the aligned magnetic field configuration, as can be confirmed with our numerical results presented in the following section.

\section{Numerical Results}

Whether the reverse shock is relativistic or not is mainly depending on the dimensionless parameter $\xi$, which is defined as $\xi=(l/\Delta_0)^{1/2}\eta^{-4/3}$ if the circum-burst environment is ISM (Sari \& Piran 1995). The Sedov length $l$ depends on the isotropic kinetic energy of the GRB ejecta and the density of ISM, i.e., $l=(E_{iso}/n_1m_pc^2)^{1/3}$. If the shell width $\Delta_0$ is large enough (thick shell case), $\xi<1$ and the reverse shock is relativistic (RRS). On the contrary, if the shell width $\Delta_0$ is small enough (thin shell case), $\xi>1$ and the reverse shock is initially non-relativistic (NRS). According to the above different hydrodynamics (RRS or NRS) and which shocked region (Region 2 or Region 3) dominates the optical emission at early times, we explore the following four cases: (1) thick shell + reverse shock dominated, (2) thin shell + forward shock dominated, (3) thick shell + forward shock dominated, and (4) thin shell + reverse shock dominated.

Based on the equations in Section 2, we numerically calculate the dynamical evolution of the system. In the thick shell case, we take the parameter values as follows: $E_{52}=E_{iso}/10^{52}\,{\rm erg}=1$, $n_1=1\, {\rm cm}^{-3}$, $\eta=300$, $\theta_j=0.1$, and $\Delta_{0,12}=\Delta_0/10^{12}\,{\rm cm}=3$. In the thin shell case, we adopt the parameter values as: $E_{52}=0.01$, $n_1=1\,{\rm cm}^{-3}$, $\eta=100$, $\theta_j=0.1$, and $\Delta_{0,12}=0.03$. We assume an adiabatic shock, i.e., the radiation coefficient $\varepsilon_{rs}=\varepsilon_{fs}=0$. The evolution of the bulk Lorentz factor of the shocked region are shown in Fig. 1 for the thin shell case and in Fig. 2 for the thick shell case. The main difference between the dynamics given by HDL99 and ours appears around the reverse shock crossing time $t_c$.

We numerically calculate the polarization evolution of an early optical afterglow (R-band). The equal arrival time surface effect and the lateral expansion are not considered in our calculation. The dynamical parameters in Cases 1 and 3 are the same as listed in the above paragraph for the thick shell case. The dynamical parameters in Cases 2 and 4 are the same for the thin shell case. The shock microphysics parameters are as follows: (1) $\varepsilon_{e,rs}=\varepsilon_{B,rs}=0.1$, $\varepsilon_{e,fs}=0.05$, $\varepsilon_{B,fs}=0.002$ for Case 1; $\varepsilon_{e,rs}=0.015$, $\varepsilon_{B,rs}=0.01$, $\varepsilon_{e,fs}=0.02$, $\varepsilon_{B,fs}=0.005$ for Case 2; $\varepsilon_{e,rs}=0.01$, $\varepsilon_{B,rs}=0.005$, $\varepsilon_{e,fs}=0.02$, $\varepsilon_{B,fs}=0.01$ for Case 3; $\varepsilon_{e,rs}=\varepsilon_{B,rs}=0.1$, $\varepsilon_{e,fs}=0.05$, $\varepsilon_{B,fs}=0.002$ for Case 4. A fraction $\varepsilon_{e,rs}$ of the internal energy in the reverse shock region goes into the electrons. A fraction $\varepsilon_{e,fs}$ and $\varepsilon_{B,fs}$ of the internal energy in the forward shock region go into the electrons and random magnetic field, respectively. The power law index of the energy distribution ($N(\gamma_e)\propto\gamma_e^{-p}$) for the shock heated electrons is $p_{rs}$ for the reverse shock and $p_{fs}$ for the forward shock, and $p_{rs}=p_{fs}=2.5$ is assumed. We let the linear polarization degree of synchrotron radiation in ordered magnetic field as $\pi_0=0.6$ in R-band. The orientation of the aligned magnetic field is assumed to be $\delta=\pi/4$. In our calculation, we assume that the GRB is located at redshift $z=1$, and adopt a flat Universe with $\Omega_M=0.27$, $\Omega_\Lambda=0.73$ and $H_0=71{\rm \,km\,s^{-1}\,Mpc^{-1}}$.

For each case, we discuss two magnetic field configurations. Figs. 3 and 4 show the light curves and polarization evolution of optical afterglows with different magnetic field configuration in Case 1. The magnetic field is toroidal for Fig. 3  and aligned for Fig. 4. The reverse shock crossing time in these two figures is 123 s. In Fig. 3, the flux of the reverse shock emission peaks at $t_c$. When $q\equiv\theta_V/\theta_j=0.0$, the LOS locates at the center of the jet, and the polarization degree $P_{60}=\Pi_T$ is zero because of the axis-symmetry. When $q=0.6$ and $1.0$, the magnetic field before $t_c$ can be regarded as large-scale aligned within the observable $1/\gamma$ cone centered around the LOS. Therefore, the polarization degree almost keeps as a constant and reaches about 45\% during the reverse shock crossing the ejecta, as can be seen in Fig. 3. After $t_c$, as the bulk Lorentz factor decreases, the observable cone increases and more and more complete toroidal field contributes to the observed polarized emission, leading to slow decrease of the net linear polarization degree. The decrease of the polarization degree is also mainly caused by the increasingly dominant forward shock emission at late times. For $q=2.0$ and $3.0$, the LOS locates outside of the jet cone, a rapid increase of the flux will appear. Our choice of model parameters make the forward shock emission comparable to the reverse shock emission around $t_c$ for $q=2.0$ and $3.0$, which causes the net polarization degree for these two observing angles about 10\% at $t_c$. A 90$^{\circ}$ rotation of the polarization angle is expected when the polarization degree crosses over zero in Fig. 3. In Fig. 4, due to the nonzero $U_{\nu,3}$ in the aligned magnetic field configuration, the evolution of the position angle is much complicated, which can be estimated by Eq.(28). The position angles for almost all the observing angles (except $q=3.0$) change their direction by $90^\circ$ abruptly once or twice with nonzero polarization degree, and such changes happen around the crossing time $t_c$. As interpreted from the theoretical aspect in the previous section, this abrupt $90^{\circ}$ change of the position angle corresponds to the moment when $Q_{\nu}$ crosses zero due to the transition of the dominant term by $Q_{\nu,2}$ and $Q_{\nu,3}$ which have different signs. As long as the LOS locates within the jet cone ($q=0.0,0.6$, and $1.0$), the polarization degree is always at a constant high level of $P_{60}=\Pi_A\sim50\%$ before the reverse shock crosses the ejecta.

Figs. 5 and 6 correspond to Case 2. The crossing time of the reverse shock in this case is about 66 s. The magnetic field configuration is toroidal for Fig. 5 and aligned for Fig. 6. As the forward shock dominates the whole emission, the net polarization degree for $q\leq1$ (LOS within the jet cone) is nearly zero or very small for the whole time. For $q=2.0$ and $3.0$, as the LOS locates far from the jet cone, the polarization degree at late time ($t>t_c$) can reach a relatively high level ($\sim50\%$). However, such polarized emission is difficult to be detected, because prompt GRB emission should be extremely weak and can not trigger the gamma-ray detector for such a large viewing angle. In Fig. 6, the position angle for $q=0.0$ stays constant for the whole time, while the change of the position angle for the other $q$ values is gradual. It should be noted that according to the lower panel of Fig. 6, the position angle for $q=$1.0, 2.0, and 3.0 is the same for the early and late times, and its maximal change ($\leq\pi/4$) happens around the reverse shock crossing time $t_c$.

Figs. 7 and  8 show the light curves and polarization evolution of optical afterglows with different magnetic field configuration in Case 3. The crossing time of the reverse shock in this case is about 123 s. Fig. 7 corresponds to a toroidal magnetic field configuration, while Fig. 8 corresponds to an aligned magnetic field configuration. As we know, the difference between Case 2 and Case 3 is that in the former case the reverse shock is initially non-relativistic (thin shell) while in the latter case the reverse shock is relativistic (thick thell). So the rising part of the light curve before the afterglow peak has different power law slope for these two cases. Another main difference between Case 2 and Case 3 is the polarization position angle evolution in the aligned magnetic field configuration. Contrary to Case 2 (lower panel in Fig. 6), the position angle is different for different viewing angles (i.e. for $q=1.0$, 2.0, and 3.0) at early times ($t<t_c$), as can be seen in Fig. 8. For viewing angles $q=0.0$ and 0.6, the position angle before the reverse shock crossing time is almost a constant. All of the position angles with different $\theta_V$, expect for $\theta_V=0.0$, are convergent to the same value soon after the reverse shock crossing time.

Figs. 9 and 10 correspond to Case 4. The crossing time of the reverse shock is about 66 s. The magnetic field configuration is toroidal for Fig. 9 and aligned for Fig. 10. In Case 4, for both magnetic field configurations, there is a peak in the polarization degree evolution at $t_c$ for all the observing angles discussed in this paper, expect that for $q=0$ in the toroidal magnetic field configuration. In the toroidal magnetic field configuration, the peak value of $\Pi_T$ is quite different, $\Pi_T\sim 40\%$ for $\theta_V\leq\theta_j$, while $\Pi_T$ is much smaller for $\theta_V>\theta_j$. After the crossing time the polarization degree declines and crosses over zero, which corresponds to a sudden rotation of the position angle by 90$^{\circ}$. In the aligned magnetic field configuration, the peak value of $\Pi_A$ is $\sim 30\% - 40\%$ no matter the LOS locates inside the jet cone or outside the jet cone. In this configuration, the position angles for almost all the observing angles (except $q=3.0$) change their direction by $90^\circ$ abruptly twice, once before $t_c$ and another after $t_c$. For the aligned magnetic field configuration, when position angle changes its direction abruptly by $90^\circ$, the polarization degree is not necessarily to be zero. Although Case 4 and Case 1 are both reverse shock dominated cases, the afterglow light curves have different rising slopes before $t_c$. Typically, Case 4 (thin shell) has more rapid rising light curve than Case 1 (thick shell). According to our results, the polarization evolution for these two cases are also different. Before the reverse shock crossing time, the polarization evolution in Case 4 is more quick than that in Case 1.

In general, we find that in Cases 1 and 4, i.e. the early afterglow is dominated by the reverse shock, the position angle changes abruptly by $90^\circ$ at the vicinity of the crossing time ($0.1t_c<t<10t_c$). Furthermore, if the polarization angle changes by $90^\circ$ twice, or such a change happens before $t_c$, or the position angle changes gradually, then we can identify that the magnetic field configuration in the ejecta might be aligned. Otherwise, the magnetic field configuration may be toroidal. In Cases 2 and 3, i.e. the forward shock dominates the whole emission, no matter what the magnetic field configuration is in the reverse shocked region, the polarization degree is very small and nearly zero for all observing angles before the reverse shock crossing time. For all the four cases discussed above, at large observing angles (LOS outside the jet cone), e.g. $q=2.0$ and $3.0$, there is a peak in the polarization degree evolution curve at late times (about $10^4-10^6$ s) in our calculations. The peak time has a correlation with the observing angle. The larger the observing angle, the later the peak time. For the toroidal magnetic field configuration in Region 3, if the observing angle is zero, because of the symmetry in this magnetic field configuration, the polarization degree is always zero, even the emission is dominated by the reverse shock. The jet break time (when $\gamma \theta_j\sim1$) is about $4\times10^3$ s in the thin shell case and about $2\times10^4$ s in the thick shell case in our calculations. When $q=0.6$, for the toroidal magnetic field configuration, our results show that the position angle changes abruptly by $90^\circ$ slightly after jet break time for the four cases. For the aligned magnetic field configuration, the position angle changes abruptly by $80^\circ$ slightly after jet break time for the reverse shock dominated cases (Cases 1 and 4). However, the change of the position angle slightly after the jet break time for the forward shock dominated cases (Cases 2 and 3) is very small (about $6^\circ$). The reason is that the reverse shock emission in our calculation also affects the Stokes parameter values at late times, even its contribution to the total flux can be neglected.

\section{Application to GRB 120308A}
There have been few early optical afterglows with polarized emission detected in the past decade. These GRB afterglows include GRB 060418 with an upper limit $\Pi<8\%$ during the forward shock dominated phase (Mundell et al. 2007), GRB 090102 with an averaged linear polarization $\Pi\sim10\%$ during the reverse shock dominated phase (Steele et al. 2009), GRB 120308A with an evolving polarization degree and a nearly constant position angle from the reverse shock emission phase to the forward shock emission phase (Mundell et al. 2013).

In this Section we apply our models to try to interpret the afterglow observations of GRB 120308A. Our fitting results are shown in Fig. 11 for a toroidal magnetic field configuration and in Fig. 12 for an aligned magnetic field configuration. Both of the magnetic field configurations can fit the observations equally well. The dynamics parameters we use both for totoidal and aligned magnetic field configurations are: $\theta_j=0.015$, $\Delta_0=10^{11}\,{\rm cm}$, $E_{iso}=5\times10^{53}\,{\rm ergs}$, $\eta=350$, $n_1=0.01\,{\rm cm}^{-3}$, $\varepsilon_{rs}=\varepsilon_{fs}=0$. We also take $\pi_0=0.6$ for both magnetic field configurations. The other parameters adopted in our fittings are $p_{rs}=2.15$, $p_{fs}=2.1$, $\varepsilon_{e,rs}=0.044$, $\varepsilon_{B,rs}=0.1$, $\varepsilon_{e,fs}=0.05$, $\varepsilon_{B,fs}=0.018$, and $q=0.7$ for the toroidal magnetic field configuration, while $p_{rs}=2.15$, $p_{fs}=2.06$, $\varepsilon_{e,rs}=0.043$, $\varepsilon_{B,rs}=0.1$, $\varepsilon_{e,fs}=0.072$, $\varepsilon_{B,fs}=0.015$, $q=0.0$, and $\delta=\pi/6$ for the aligned magnetic field configuration. The redshift of GRB 120308A is 2.2 (Mundell et al. 2013). For the two magnetic field configurations, all the fitting parameters we use are in a reasonable range compared with typical values of GRB model parameters. When fitting the light curve and polarization evolution of GRB 120308A, for simplicity we just use the mean frequency at the $R$-band and do not integrate over the frequency band from 555 nm to 690 nm at which the observations were carried out. The difference of the polarization degree for the integration over energy band (555-690 nm) and single R-band will not exceed 3\%.

Due to the sparse data and large error bars of the polarization observations, we currently cannot distinguish between these two magnetic field configurations for GRB 120308A. Long-time observations with high polarization resolution are needed in future to nail down the true magnetic field configuration, especially observations of the polarization degree when the position angle changes abruptly by $90^\circ$ during the early afterglow phase.

\section{Conclusions and Discussion}
In this paper, we have developed the forward-reverse shock dynamics to calculate the light curve and polarization evolution of early optical afterglows of GRBs. As an example, the observed data of GRB 120308A have been fitted using our model.

The difference between our dynamics and that of HDL99 is that we take into account the reverse shock process while HDL99 neglected the effect of the reverse shock on the jet hydrodynamical evolution. Our numerical results show that the main difference is before the reverse shock crossing time $t_c$. At the beginning, the reverse shock is very weak. So the bulk Lorentz factor of the shocked region in the two dynamical models show negligible difference. In the following time, a non-negligible (even large) fraction of the bulk motion of the ejecta goes into the reverse shocked region. So our bulk Lorentz factor of the shocked region is smaller than that of HDL99 at the same observer time. After the reverse shock crosses the ejecta, Region 3 begins to expand adiabatically. The work done by Region 3 to Region 2 makes the bulk Lorentz factor of the shocked region decreases slowly. This is the reason for that our bulk Lorentz factor is a bit larger than that of HDL99 after $t_c$.

The early optical afterglow light curve depends on which shock dominates the emission, the reverse shock or the forward shock. The hydrodynamics of the reverse shock can be divided into two cases, the thick shell case which corresponds to a relativistic reverse shock, and the thin shell case which corresponds to an initially non-relativistic reverse shock. Therefore, there are four situations in total for early optical afterglows, such as the thick-shell reverse shock dominated afterglows (Case 1), the thin-shell forward shock dominated afterglows (Case 2), the thick-shell forward shock dominated afterglows (Case 3), and the thin-shell reverse shock dominated afterglows (Case 4).

While the magnetic field in the forward shock is random magnetic field, in the reverse shocked region, the magnetic field can be large-scale ordered. This ordered magnetic field is initially in the GRB ejecta carried out from the central engine. Magnetic dissipations during the prompt GRB phase may reduce the magnetization degree to a low level so that the ejecta is dominated by baryons and leptons in the afterglow phase but the large-scale structure of the magnetic field is remained. Polarization evolution of early afterglows are mainly determined by the detailed magnetic field configurations. Motivated by the main central engine mechanisms, such as magnetar dipole radiation/wind or the Blandford-Znajek process (Blandford \& Znajek 1977), we consider two magnetic field configurations, i.e., toroidal and aligned. For simplicity, all magnetic fields are assumed to be in the shock plane (see also Toma et al. 2009).

If the emission is dominated by the forward shock (Cases 2 and 3), the polarization degree is very small for all the observing angles before the reverse shock crossing time. If the emission is dominated by the reverse shock (Cases 1 and 4) at early stage, the polarization degree before $t_c$ is $\sim30\% - 50\%$ and does not change with time in the thick shell case (Case 1), while it increases to a maximal value of $\sim30\% - 50\%$ at the crossing time in the thin shell case (Case 4). The above statement does not apply to the case of $\theta_V\sim0$ with a toroidal magnetic field configuration, because the axis-symmetry in this case would lead to totally unpolarized emission. For the four cases discussed in this paper, when the observing angle is larger than the half opening angle of the jet, which means that the LOS locates outside of the jet cone, there will be a peak in the polarization degree evolution at late times ($t>t_c$) and the peak time has a positive correlation with the observing angle (as can be seen in the figures for $q=2.0$ and $3.0$).

For the aligned magnetic field configuration, the position angle in Case 4 changes abruptly by $90^\circ$ twice around (before and after) the crossing time $t_c$ for all the observing angles (except $q=3.0$). The position angle in Case 1 changes abruptly by $90^\circ$ once soon after the crossing time, for all the observing angles (expect $q=3.0$). When $q=0.6$, the position angle changes abruptly by $90^\circ$ slightly after the jet break time in all of the four cases with the toroidal magnetic field configuration. For the aligned magnetic field configuration, when $q=0.6$, the position angle changes abruptly by $80^\circ$ slightly after jet break time in Cases 1 and 4. For comparison, in Cases 2 and 3, the change of the position angle slightly after the jet break time is $\sim 6^\circ$. These results are generally consistent with that of Sari (1999).

With polarization observations during some GRB early afterglow phase, we can distinguish between magnetic field configurations for these GRBs. For example, in Cases 1 and 4, the position angle changes abruptly by $90^\circ$ at the vicinity of the crossing time ($0.1t_c<t<10t_c$). If such a polarization angle change happens twice, or such a change happens before $t_c$, or the position angle changes continuously, then we can identify that the magnetic field configuration in the ejecta might be aligned. Otherwise, the magnetic field configuration may be toroidal. It should be emphasized that if we observe an abrupt $90^{\circ}$ change in the position angle with a nonzero polarization degree, then we can infer the magnetic field configuration to be aligned. Otherwise, if we observe an abrupt $90^{\circ}$ change in the position angle with a zero polarization degree, then the magnetic field configuration should be toroidal. The magnetic field configuration is essential for us to understand the GRB central engine. If the field configuration in the ejecta is aligned, the possible central engine is a magnetar. If the field configuration in the ejecta is toroidal, the probable central engine is a black hole.

We applied our model to the observed data of GRB 120308A, and found that both of the magnetic field configurations can fit the data equally well. The parameters we use for both of the magnetic field configurations are all reasonable compared with other GRB modellings. Currently we can not distinguish between the two magnetic field configurations for GRB 120308A. However, future polarization observations with higher quality and more data, especially the polarization degree observations when the position angle changes abruptly by $90^\circ$ or the observations of the gradual evolution of the position angle, are encouraged to pin down the true magnetic field configuration and provide important clues on the GRB central engines.

The early afterglows arising from reverse-forward shocks are commonly observed at optical band. Observations of the early X-ray emission from the external reverse shock are extremely rare, because according to the standard GRB afterglow model, the forward shock emission dominates over the reverse shock emission at X-ray band. Moreover, early X-ray afterglows from the external shocks are usually hidden by the high latitude emission of the prompt GRB or the subsequent X-ray flares. There are a good fraction of GRBs observed with the so-called X-ray shallow decay phase or X-ray plateaus, of which the timescale is long enough (typically lasts for 10 ks) and the X-ray flux is high enough for the polarization observation. Several scenarios were proposed to interpret these X-ray plateaus, such as the matter-dominated energy injection model and Poynting-flux dominated energy injection model for the external plateaus, and the magnetic dissipation model for the internal plateaus. The latter two predict relatively high linear polarization for X-ray plateaus, while the matter-dominated energy injection model does not. So polarization observations in the GRB X-ray plateaus can be used to constrain these models and a detailed discussion will be presented elsewhere.

\acknowledgements
We thank Y. F. Huang for useful discussions. This work was supported by the National Basic Research Program (``973'' Program) of China (grant Nos. 2014CB845800 and 2013CB834900) and the National Natural Science Foundation of China (grant Nos. 11573014 and 11322328). X.F.W was also partially supported by the One-Hundred-Talent Program, the Youth Innovation Promotion Association (2011231), and the Strategic Priority Research Program ``The Emergence of Cosmological Structure'' (grant No. XDB09000000) of the Chinese Academy of Sciences, and the Natural Science Foundation of Jiangsu Province (No. BK2012890).

\begin{figure}
\includegraphics[angle=0,scale=0.8]{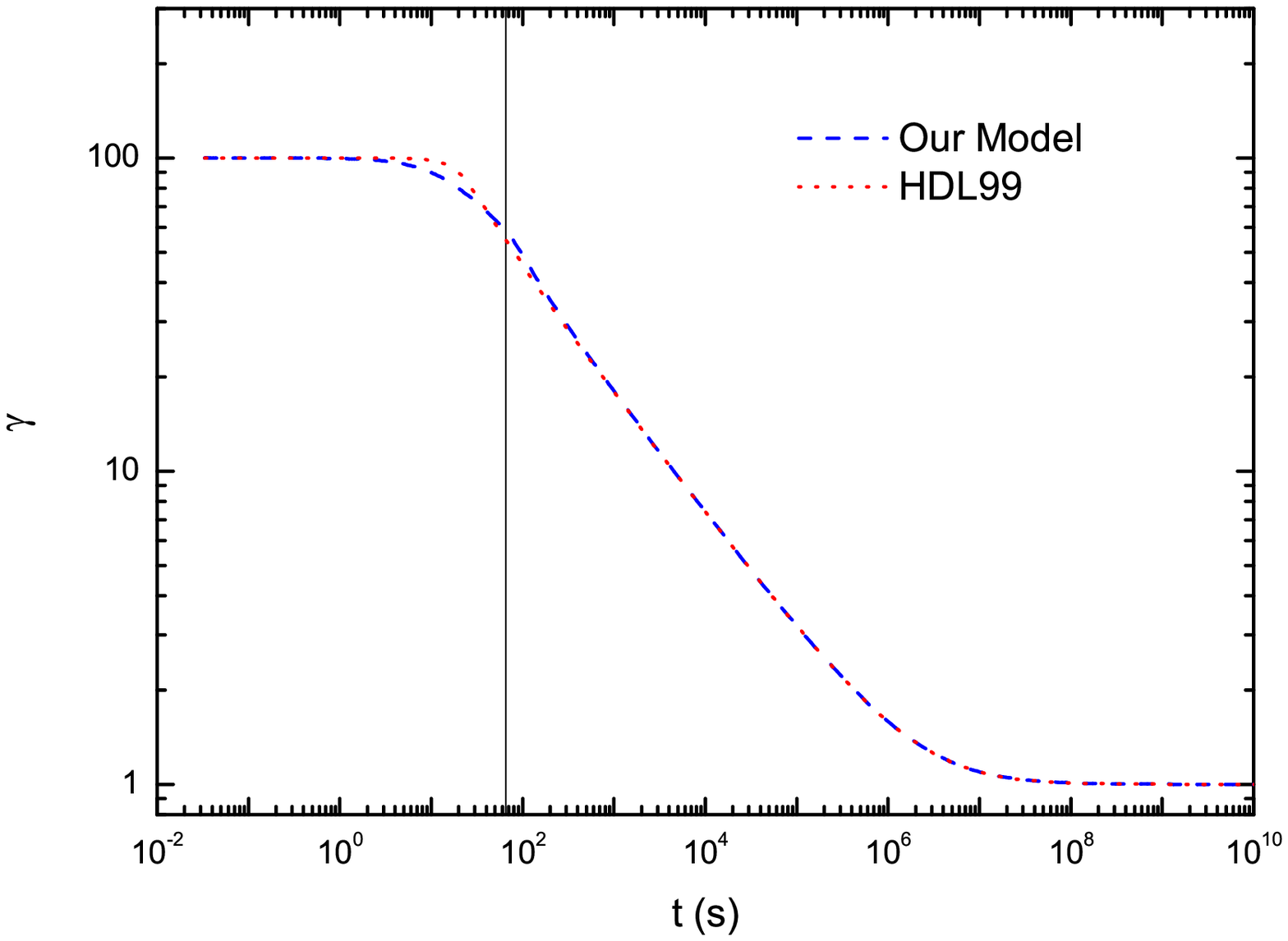}
\caption{Evolution of the bulk Lorentz factor in the thin shell case. The dashed line corresponds to our results. The dotted line shows the results of HDL99. The vertical solid line
 correspond to the reverse shock crossing time $t_c$. \label{fig1}}
\end{figure}

\begin{figure}
\includegraphics[angle=0,scale=0.8]{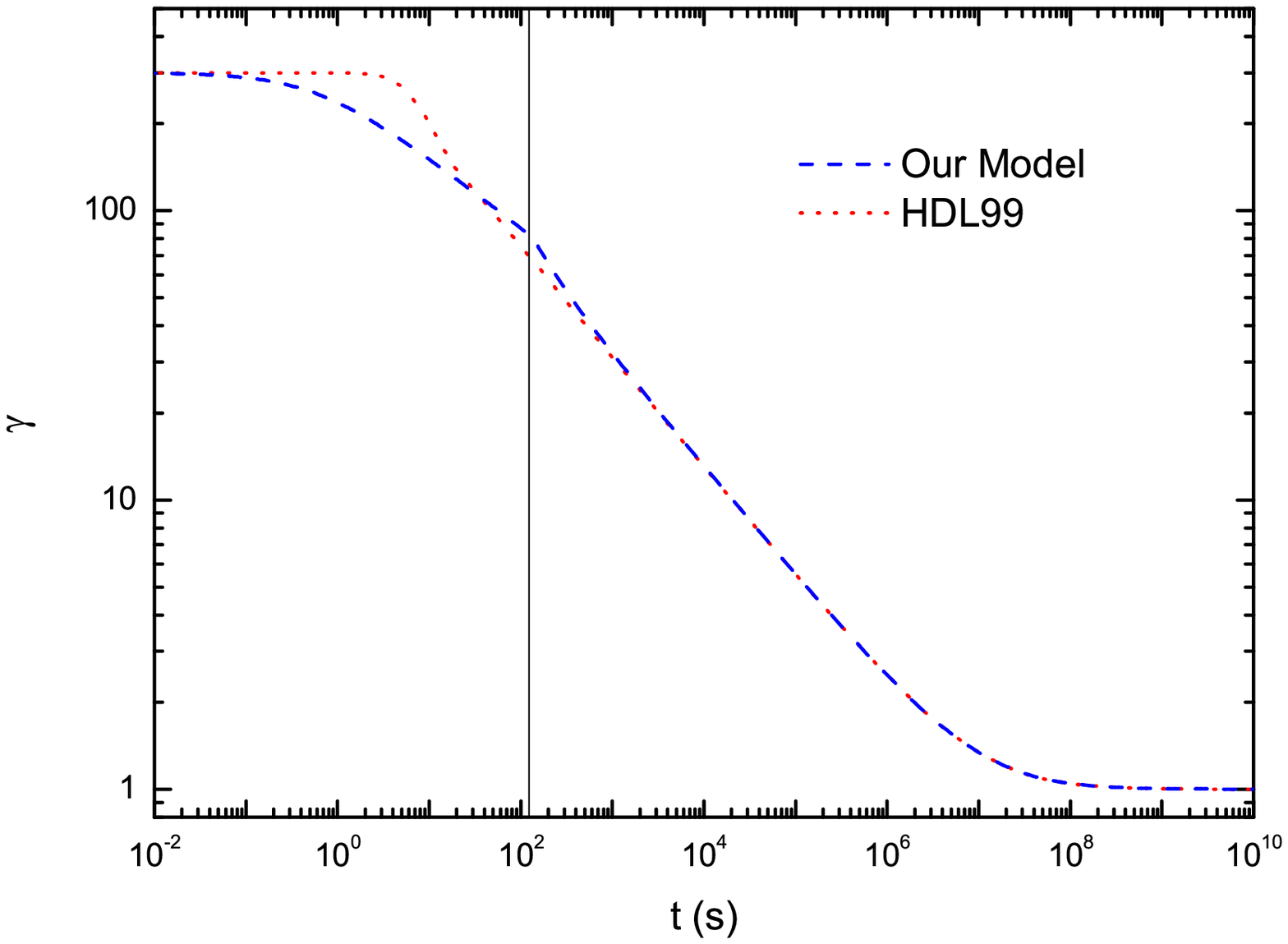}
\caption{Evolution of the bulk Lorentz factor in the thick shell case. The dashed line corresponds to our results. The dotted line shows the results of HDL99. The vertical solid
line correspond to the reverse shock crossing time $t_c$. \label{fig2}}
\end{figure}

\begin{figure}
\includegraphics[angle=0,scale=0.8]{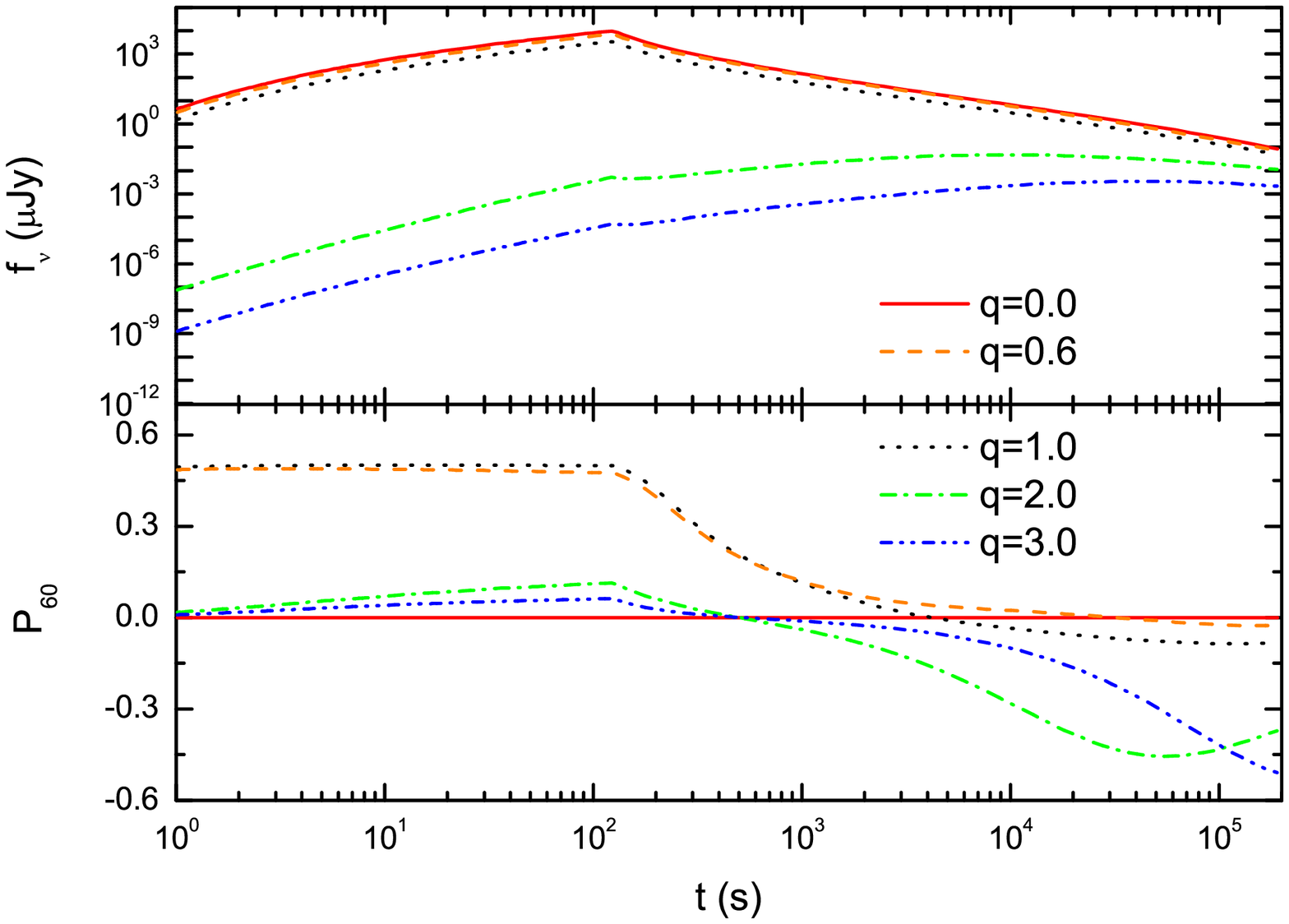}
\caption{Light curves (upper panel) and polarization evolution (lower panel) in Case 1 (i.e. thick shell + reverse shock dominated) with a toroidal magnetic field configuration. Different line styles correspond to different observing angles with $q\equiv\theta_V/\theta_j$. \label{fig3}}
\end{figure}

\begin{figure}
\includegraphics[angle=0,scale=0.8]{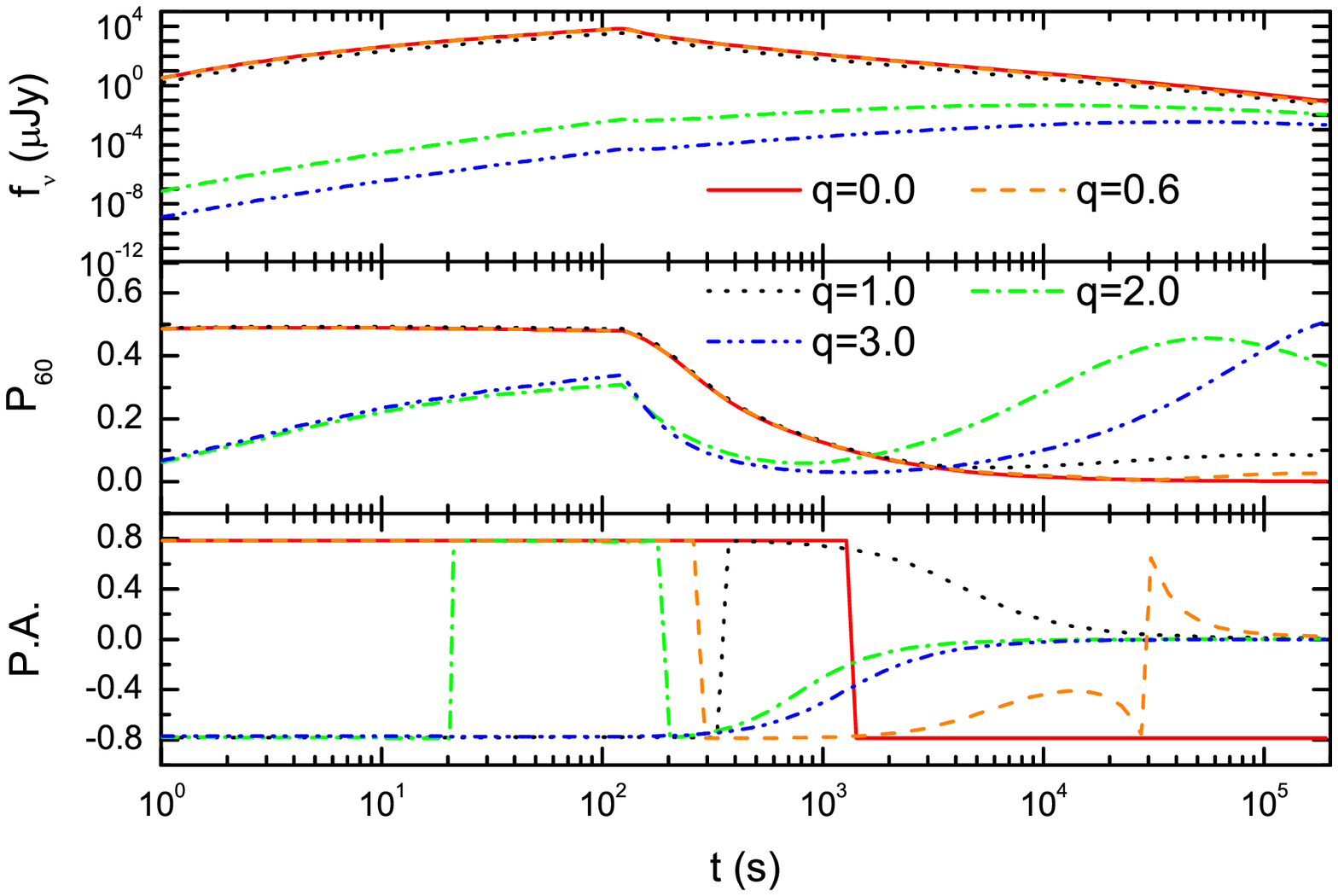}
\caption{Light curves (upper panel) and polarization evolution (middle panel) in Case 1 (i.e. thick shell + reverse shock dominated) with an aligned magnetic field configuration. The bottom panel shows the evolution of the position angle. Different line styles correspond to different observing angles with $q\equiv\theta_V/\theta_j$. \label{fig4}}
\end{figure}

\begin{figure}
\includegraphics[angle=0,scale=0.8]{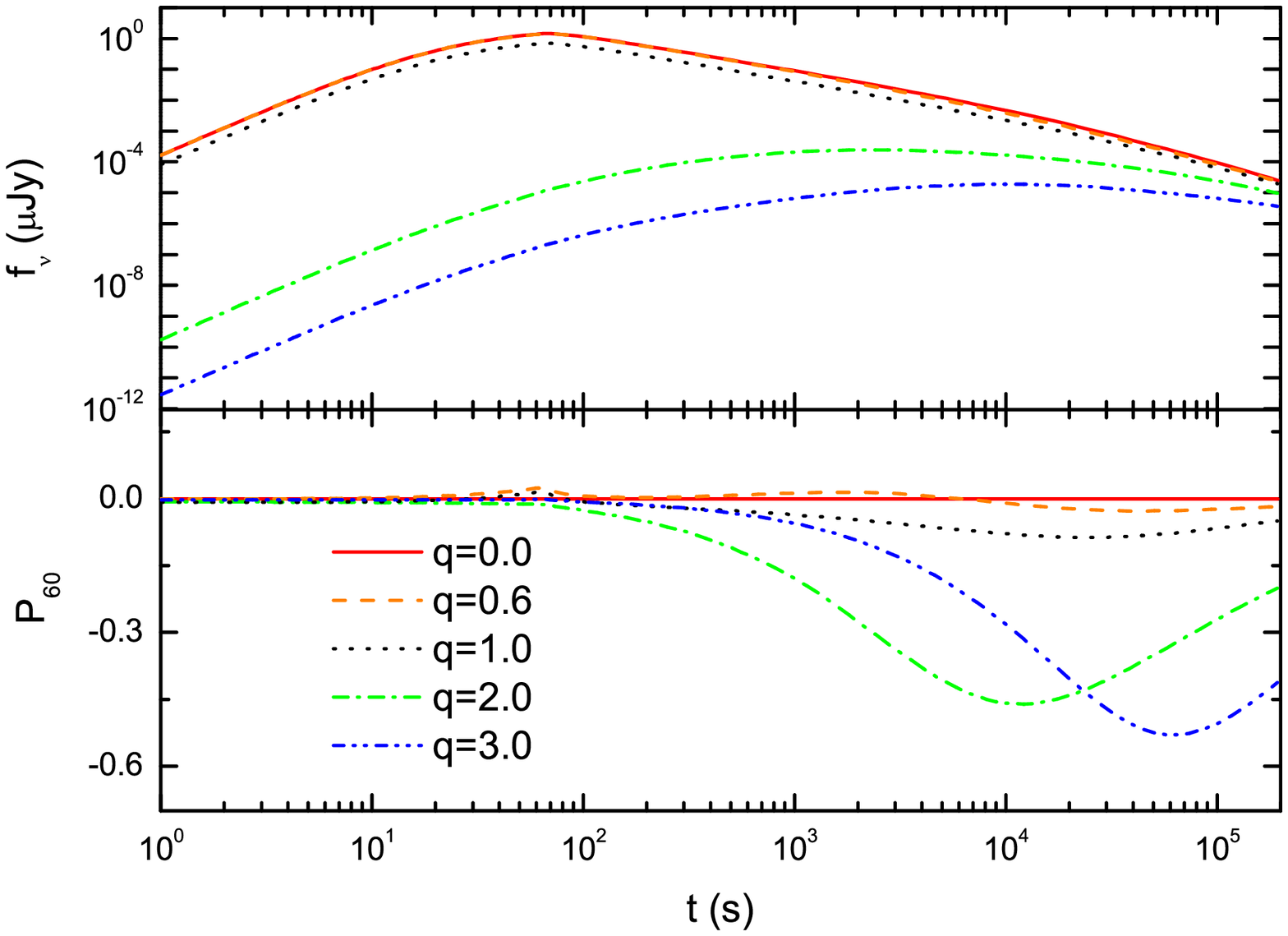}
\caption{Light curves (upper panel) and polarization evolution (lower panel) in Case 2 (i.e. thin shell + forward shock dominated) with a toroidal magnetic field configuration. Different line styles correspond to different observing angles with $q\equiv\theta_V/\theta_j$. \label{fig5}}
\end{figure}

\begin{figure}
\includegraphics[angle=0,scale=0.8]{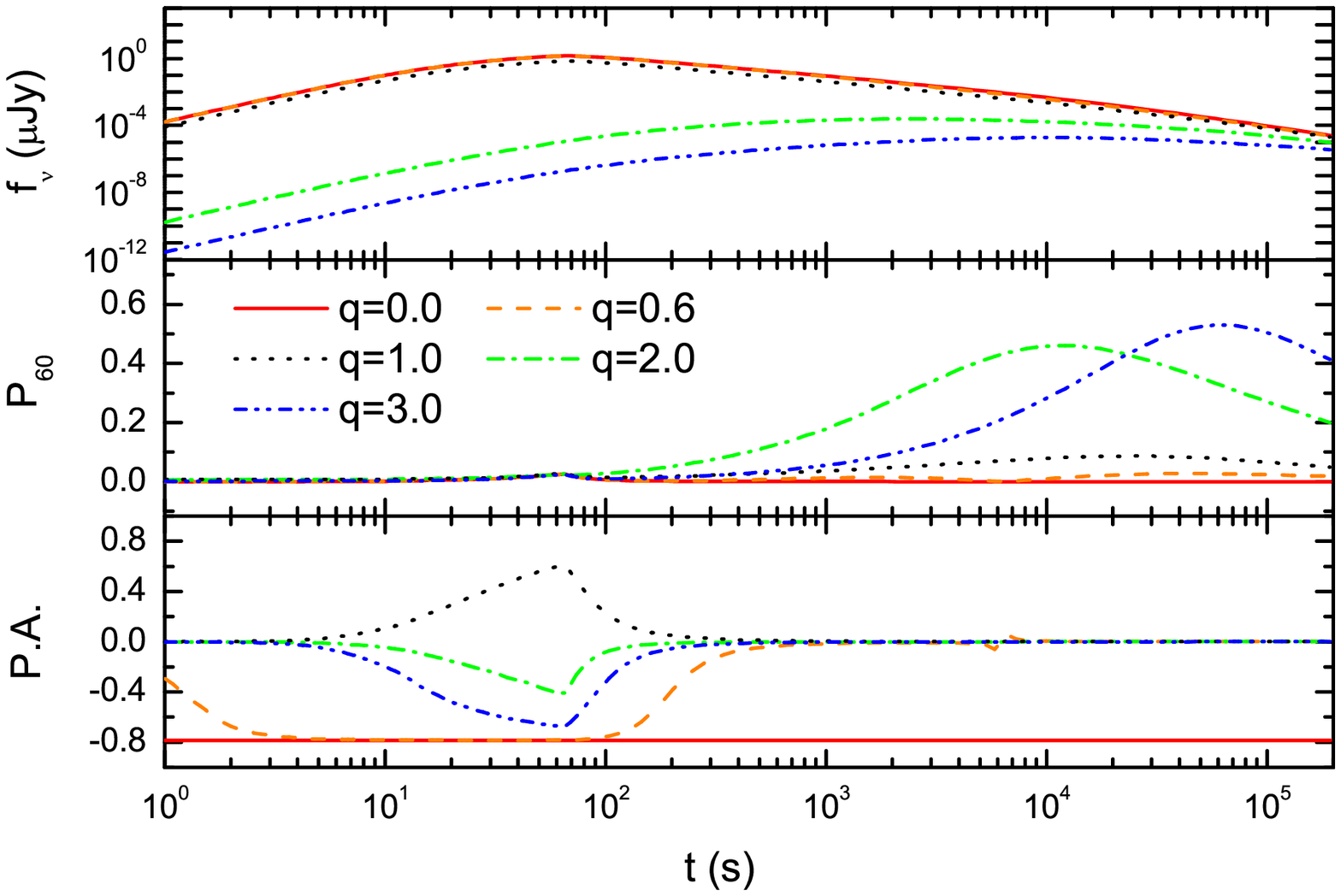}
\caption{Light curves (upper panel) and polarization evolution (middle panel) in Case 2 (i.e. thin shell + forward shock dominated) with an aligned magnetic field configuration. The bottom panel shows the evolution of the position angle. Different line styles correspond to different observing angles with $q\equiv\theta_V/\theta_j$. \label{fig6}}
\end{figure}

\begin{figure}
\includegraphics[angle=0,scale=0.8]{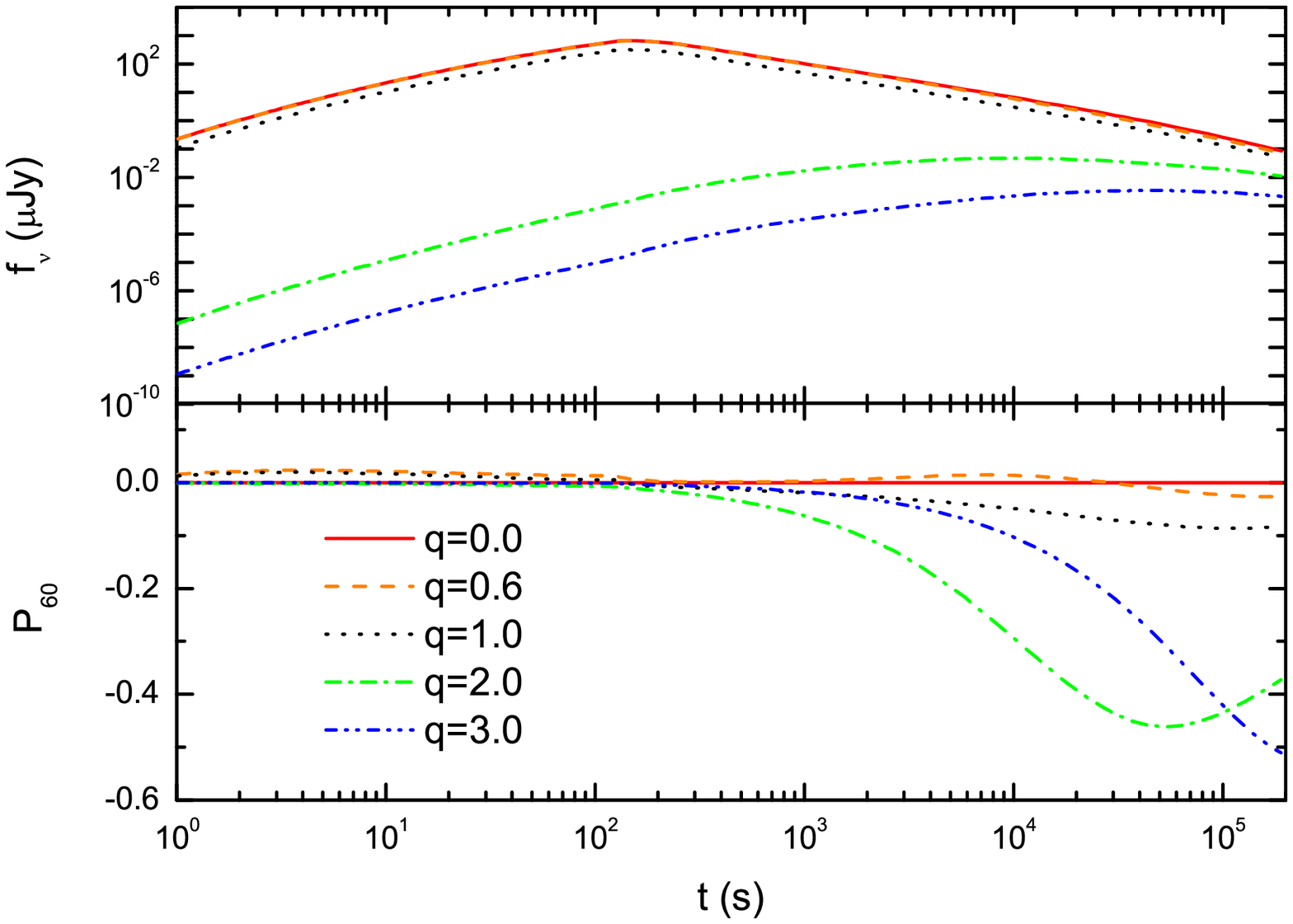}
\caption{Light curves (upper panel) and polarization evolution (lower panel) in Case 3 (i.e. thick shell + forward shock dominated) with a toroidal magnetic field configuration. Different line styles correspond to different observing angles with $q\equiv\theta_V/\theta_j$. \label{fig7}}
\end{figure}

\begin{figure}
\includegraphics[angle=0,scale=0.8]{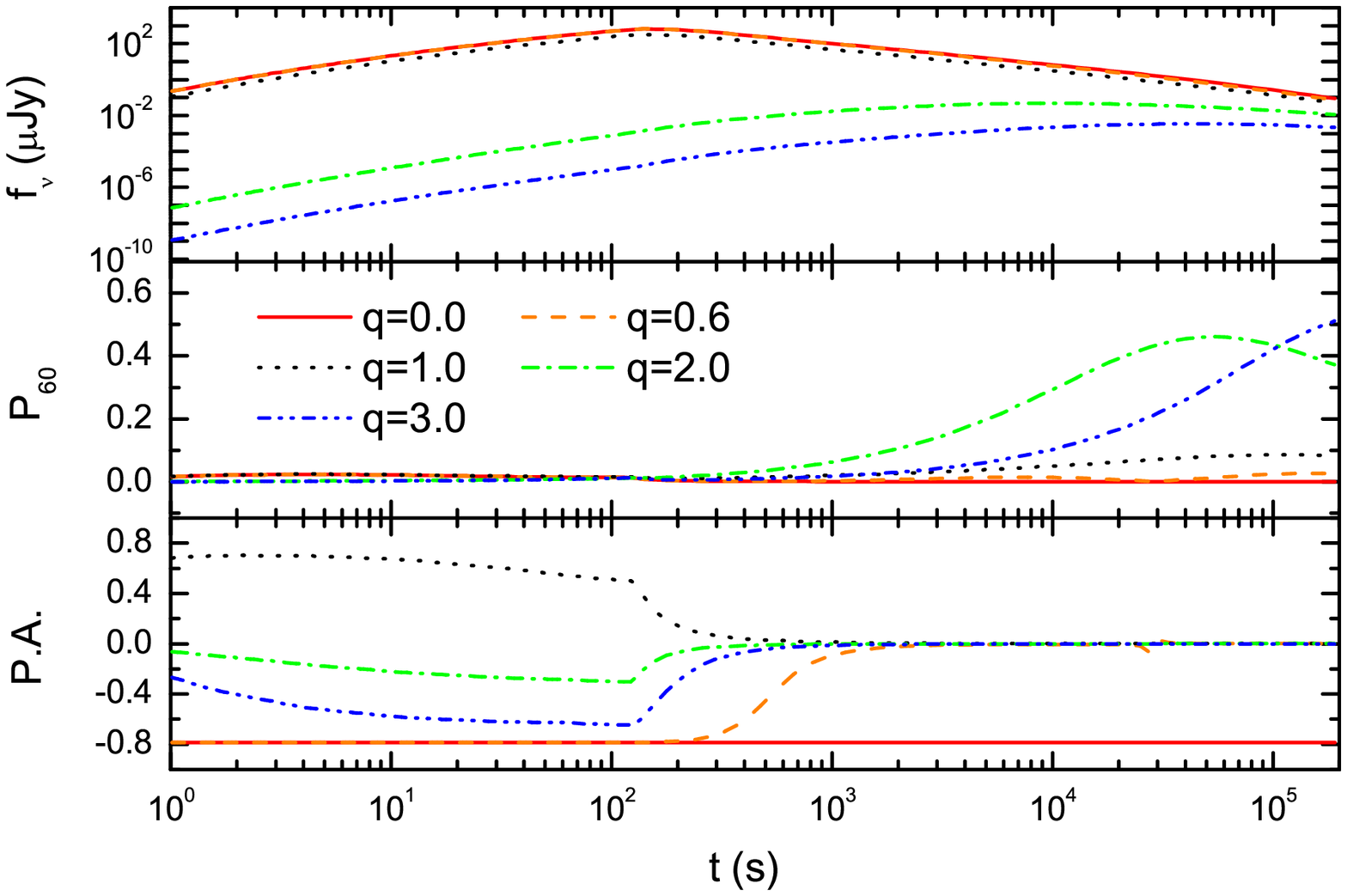}
\caption{Light curves (upper panel) and polarization evolution (middle panel) in Case 3 (i.e. thick shell + forward shock dominated) with an aligned magnetic field configuration. The bottom panel shows the evolution of the position angle. Different line styles correspond to different observing angles with $q\equiv\theta_V/\theta_j$. \label{fig8}}
\end{figure}

\begin{figure}
\includegraphics[angle=0,scale=0.8]{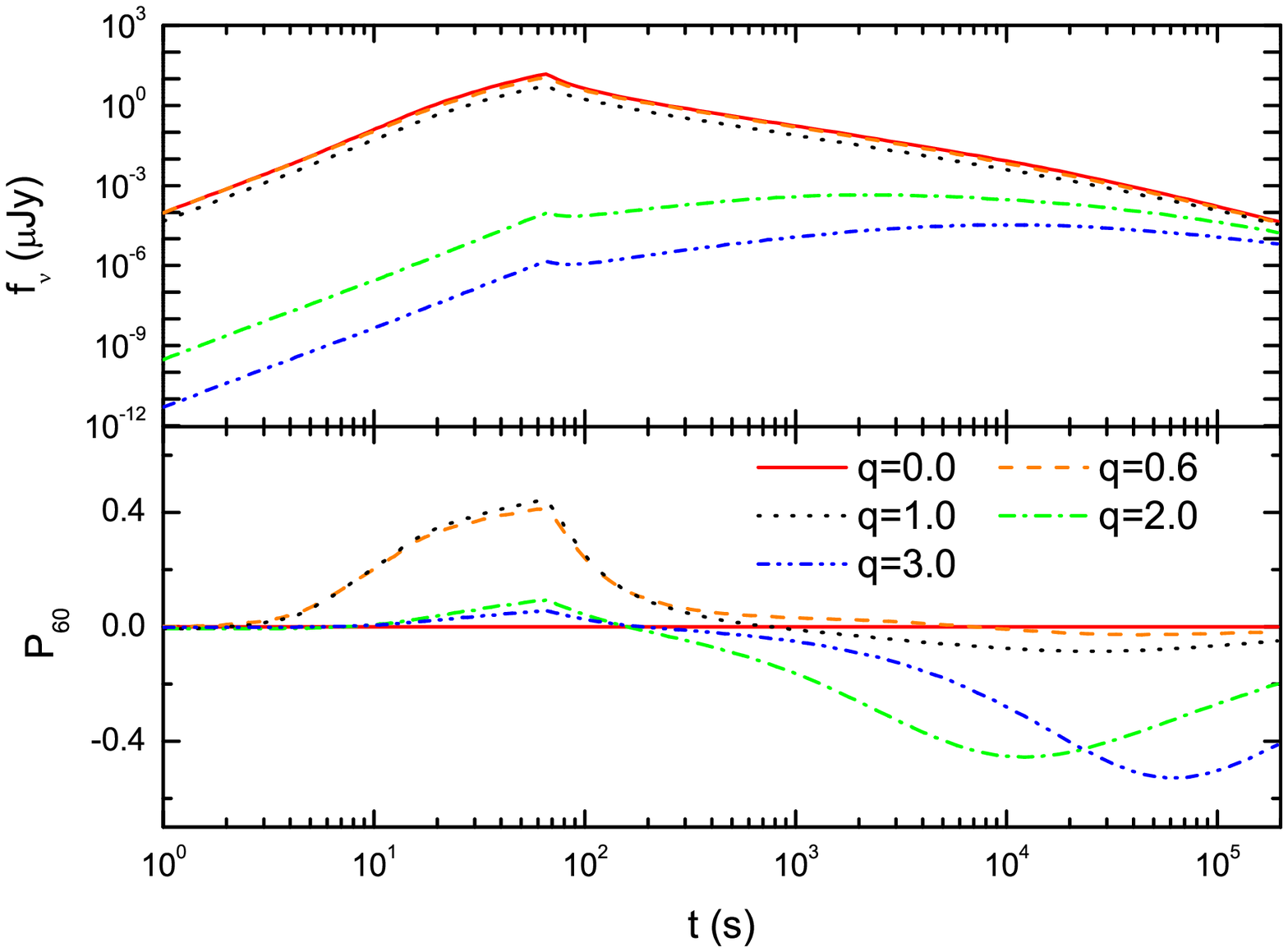}
\caption{Light curves (upper panel) and polarization evolution (lower panel) in Case 4 (i.e. thin shell + reverse shock dominated) with a toroidal magnetic field configuration. Different line styles correspond to different observing angles with $q\equiv\theta_V/\theta_j$. \label{fig9}}
\end{figure}

\begin{figure}
\includegraphics[angle=0,scale=0.8]{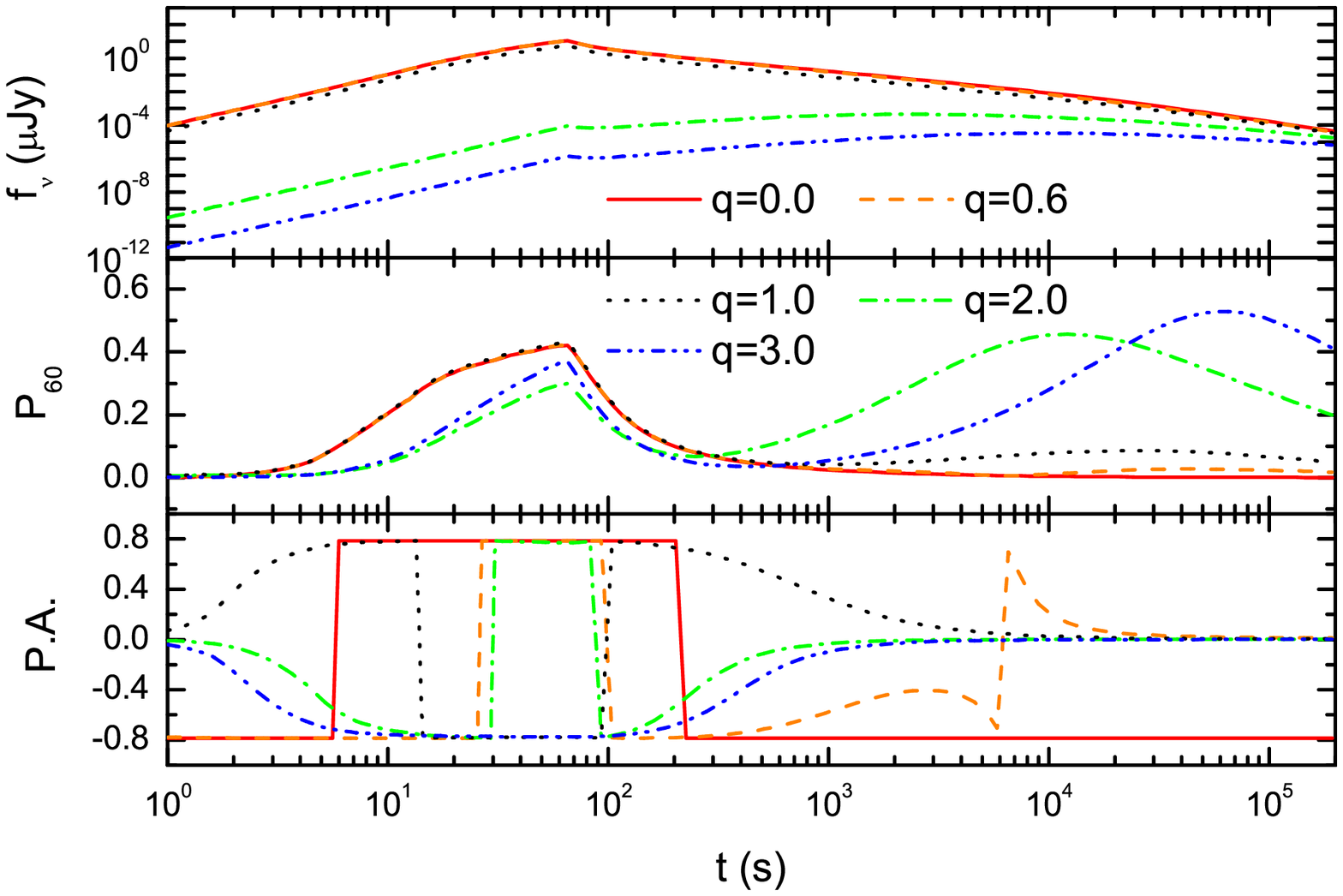}
\caption{Light curves (upper panel) and polarization evolution (middle panel) in Case 4 (i.e. thin shell + reverse shock dominated) with an aligned magnetic field configuration. The bottom panel shows the evolution of the position angle. Different line styles correspond to different observing angles with $q\equiv\theta_V/\theta_j$. \label{fig10}}
\end{figure}

\begin{figure}
\includegraphics[angle=0,scale=0.8]{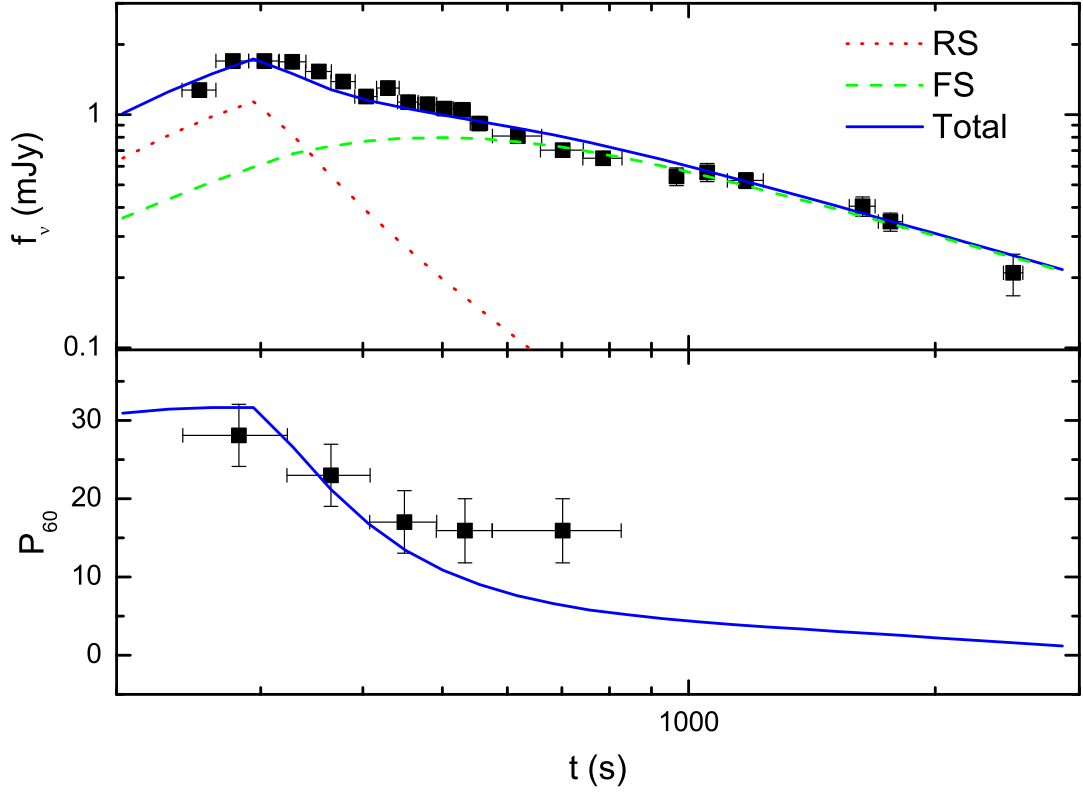}
\caption{Fitting GRB 120308A with a toroidal magnetic field configuration. The upper panel shows the evolution of the flux. The solid line is the total flux from forward-reverse shock regions. The dotted line corresponds to the flux from the reverse shocked region. The dashed line shows the flux from the forward shocked region. The lower panel shows the evolution of the polarization degree. The solid line shows our numerical result. The data are taken from Mundell et al. (2013). \label{fig11}}
\end{figure}

\begin{figure}
\includegraphics[angle=0,scale=0.8]{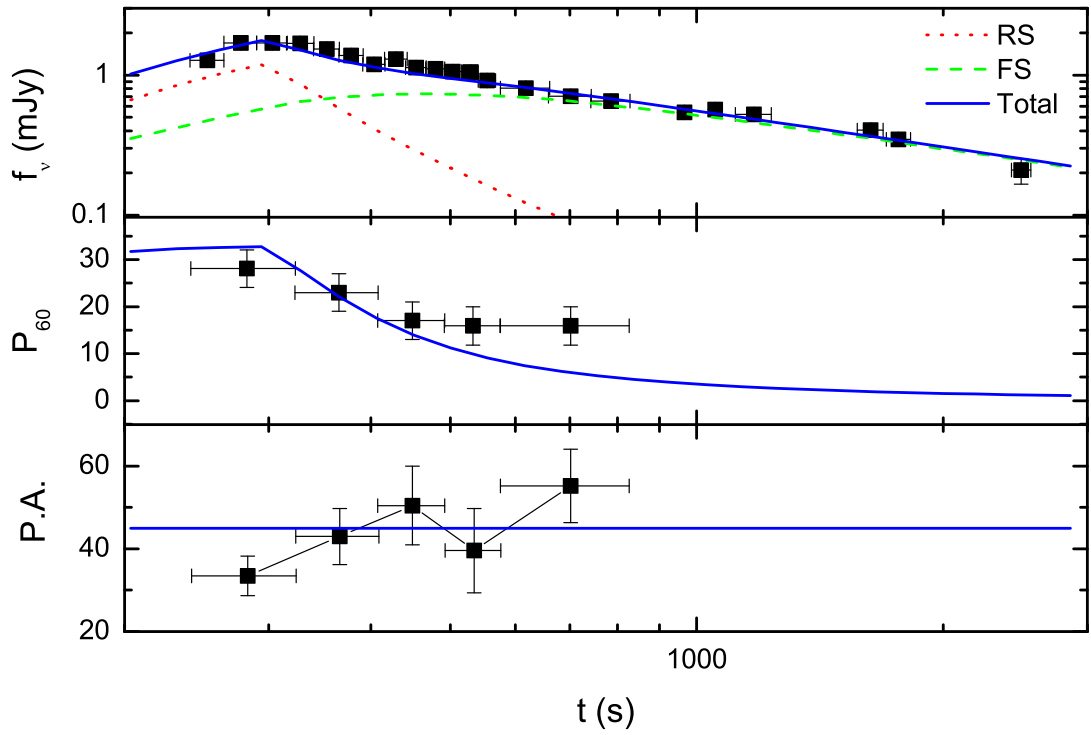}
\caption{Fitting GRB 120308A with an aligned magnetic field configuration. Same as Fig. 11 except for the lower panel, which shows the evolution of the position angle. \label{fig12}}
\end{figure}

\end{document}